\documentclass[sigconf]{acmart}
\AtBeginDocument{%
  \providecommand\BibTeX{{%
    \normalfont B\kern-0.5em{\scshape i\kern-0.25em b}\kern-0.8em\TeX}}}

\acmConference[WWW'23]{WWW'23: TheWebConf2023}{May 1--5, 2023}{Austin, TX, US}
\acmBooktitle{WWW'22: TheWebConf2023,
   May 1--5, 2023, Austin, TX, US}
\usepackage{pifont}
\newcommand{\cmark}{\ding{51}}%
\newcommand{\xmark}{\ding{55}}%
\usepackage{subfigure}
\usepackage{pgfplots}
\usepackage{tikz}
\usepackage{booktabs}
\usepackage{adjustbox}
\usepackage{multirow}
\begin{document}
%%
%% The "title" command has an optional parameter,
%% allowing the author to define a "short title" to be used in page headers.
\title{Customized Watermarking for Deep Neural Networks via Label Distribution Perturbation}

%%
%% The "author" command and its associated commands are used to define
%% the authors and their affiliations.
%% Of note is the shared affiliation of the first two authors, and the
%% "authornote" and "authornotemark" commands
%% used to denote shared contribution to the research.

\author{Tzu-Yun Chien}
\affiliation{%
  \institution{National Tsing Hua University}
  \city{Hsinchu}
  \country{Taiwan}}
\email{tfg10232338@gmail.com}

\author{Chih-Ya Shen}
\affiliation{%
  \institution{National Tsing Hua University}
  \city{Hsinchu}
  \country{Taiwan}}
\email{shenchihya@gmail.com}

%%
%% By default, the full list of authors will be used in the page
%% headers. Often, this list is too long, and will overlap
%% other information printed in the page headers. This command allows
%% the author to define a more concise list
%% of authors' names for this purpose.
\renewcommand{\shortauthors}{Tzu-Yun Chien, Chih-Ya Shen}

%%
%% The abstract is a short summary of the work to be presented in the
%% article.
\balance
\begin{abstract}
  With the increasing application value of machine learning, the intellectual property (IP) rights of deep neural networks (DNN) are getting more and more attention. With our analysis, most of the existing DNN watermarking methods can resist fine-tuning and pruning attack, but distillation attack. To address these problem, we propose a new DNN watermarking framework, \emph{Unified Soft-label Perturbation (USP)}, having a detector paired with the model to be watermarked, and \emph{Customized Soft-label Perturbation (CSP) }, embedding watermark via adding perturbation into the model output probability distribution. Experimental results show that our methods can resist all watermark removal attacks and outperform in distillation attack. Besides, we also have an excellent trade-off between the main task and watermarking that achieving 98.68\% watermark accuracy while only affecting the main task accuracy by 0.59\%.
\end{abstract}

%%
%% The code below is generated by the tool at http://dl.acm.org/ccs.cfm.
%% Please copy and paste the code instead of the example below.
%%
%\begin{CCSXML}
%<ccs2012>
% <concept>
%  <concept_id>10010520.10010553.10010562</concept_id>
%  <concept_desc>Computer systems organization~Embedded systems</concept_desc>
%  <concept_significance>500</concept_significance>
% </concept>
% <concept>
%  <concept_id>10010520.10010575.10010755</concept_id>
%  <concept_desc>Computer systems organization~Redundancy</concept_desc>
%  <concept_significance>300</concept_significance>
% </concept>
% <concept>
%  <concept_id>10010520.10010553.10010554</concept_id>
%  <concept_desc>Computer systems organization~Robotics</concept_desc>
%  <concept_significance>100</concept_significance>
% </concept>
% <concept>
%  <concept_id>10003033.10003083.10003095</concept_id>
%  <concept_desc>Networks~Network reliability</concept_desc>
%  <concept_significance>100</concept_significance>
% </concept>
%</ccs2012>
%\end{CCSXML}

%\ccsdesc[500]{Computer systems organization~Embedded systems}
%\ccsdesc[300]{Computer systems organization~Redundancy}
%\ccsdesc{Computer systems organization~Robotics}
%\ccsdesc[100]{Networks~Network reliability}

%%
%% Keywords. The author(s) should pick words that accurately describe
%% the work being presented. Separate the keywords with commas.
%\keywords{datasets, neural networks, gaze detection, text tagging}

%%
%% This command processes the author and affiliation and title
%% information and builds the first part of the formatted document.
\maketitle

\section{Introduction}

% 首先先說NN添加浮水印保障智慧財產權的用途
\noindent Deep neural network (DNN) technology has achieved great success in many fields in recent years, which brings huge commercial value. Therefore, a well-trained model can be regarded as a kind of intellectual property (IP) . 
%Just like writing, image creation, or other commodity software, the IP of DNN needs to be properly protected to preserve the interests of model owners. 
Thus, the watermarking mechanism has been applied to DNN as a means of protecting IP rights. By adding a watermark to the model, the model owner can track the model he or she owns through the watermark and prove the ownership by verifying.

% 從2017開始有人研究NN的浮水印方法,以及介紹水印方法基本要求
The DNN watermarking algorithm has been studied and proposed since around 2017~\cite{uchida2017embedding, adi2018turning, zhang2018protecting, fan2019rethinking, zhang2020passport, wang2021riga}. They mainly embedded watermarks by backdooring through a specific data set, or add designed regularizers to embed watermarks into model parameters. These studies had listed two common and basic requirements that a watermarking algorithm should have: i) \emph{Functionality-Preserving}: the watermark embedding should not impact the performance of the model too much. ii) \emph{Robustness against Removal Attacks}: after the model undergoes watermark removal attacks, the watermark should still be well preserved in the model. 

% 說明customization對浮水印方法很重要
%In addition to the above two requirements, \emph{Customization Ability} is also an important property for watermarking methods. If people apply the same watermarking algorithm to their model respectively, the watermarked models from different users will not be misidentified by each other. A good watermarking method needs to have a good customization ability to sure the ownership verification function is still available in such a situation. However, recent studies seldom do evaluations and experiments for customization ability.

% 現有方法有針對fine-tune與prune做攻擊防禦, 但是沒有對distillation做防禦研研究
Most of existing DNN watermarking methods~\cite{uchida2017embedding, adi2018turning, zhang2018protecting, fan2019rethinking, zhang2020passport, wang2021riga} used \emph{Fine-tuning attack} and \emph{Pruning attack} as removal attack. %These approaches claimed their watermarking algorithms are robust to both of them, but there are still some experimental settings not taken into account. For fine-tuning attack, they usually fine-tuned the model with a small learning rate, e.g. 0.001, while REFIT~\cite{chen2021refit} had experimented that once the learning rate is increased, the watermark of such watermarking method~\cite{uchida2017embedding} will disappear. For pruning attack, they picked the results of pruning without re-training. Although pruning causes the performance of the main task to be destroyed along with the watermark, the performance of the main task still has the opportunity of being recovered through retraining.
%Moreover,
However, \emph{Knowledge Distillation} (KD)~\cite{hinton2015distilling} should also be regarded as a kind of removal attack. 
%Knowledge distillation is a model compression technique that trains a small model to imitate a pre-trained model. 
Through KD, an adversary could replicate a model having comparable performance, without directly copying the watermarked model parameters. Nevertheless, most of the recent approaches also had not studied and demonstrated how robust they are to this kind of attack. 

Therefore, to gain a better understanding of the robustness of existing DNN watermarking methods against attacks in different settings, we conduct preliminary analysis in Section \ref{pre}. We found that the performance of watermarking of many methods will decrease when fine-tuning with higher learning rate or pruning with re-training. In addition, most of the approaches cannot resist distillation attack

% 介紹我們的方法,說明我們的方法有特別針對distillation攻擊做研究,並且有個人化能力 
To address the issues above, we propose a new customized DNN watermarking approach which is not only robust to common removal attacks (fine-tuning and pruning) but to distillation attack. Inspired by the fact that the student model imitates the output behavior of the teacher model in knowledge distillation, we proposed \emph{Unified Soft-label Perturbation (USP)} and \emph{Customized Soft-label Perturbation (CSP)} to embed the watermark into the output via perturbing the prediction probability distribution. 
The application scenario of \emph{USP} is when training a self-use model for internal use, we can use \emph{USP} to detect whether the model has been maliciously changed.
%The bias made by perturbing is regarded as the watermark signal, since the bias appearing in the output of the teacher model will probably transfer to the student model. 
For \emph{CSP}, we add a custom bit string vector to the prediction probability to cause the perturbation on the model prediction and achieve the effect of customization.
When the application scenario is we train a model and are going to authorize others to use the model, we can use \emph{CSP} to protect IP by customizing the watermark for each user.

We design a training framework having a detector paired with the model to be watermarked. The detector is a 5-layer Multi-Layer Perceptron (MLP). During the training process, the detector learns to recognize the features of the prediction probability distribution of the model.%, and update with the watermarked model simultaneously. 
After finish training,% the watermarked model, 
the owner then can use the detector to trace the model.

% 說明application scenario
%The application scenario of \emph{USP} is when training a self-use model for internal use, we can use \emph{USP} to detect whether the model has been maliciously changed.
%Since the watermark accuracy in the model is related to the model performance, we can then use the detection signature to detect whether the model has been maliciously changed. 
%When the application scenario is we train a model and are going to authorize others to use the model, we can use \emph{CSP} to protect IP. We customize the watermark for each user, when there is an authorized model breached by the user, we can use \emph{CSP} to verify the watermark, prove our ownership and find the person who breached the model.

% 我們的創新點
To our best knowledge, this is the first work that simultaneously researches defense against three types of attacks and also takes account of the tougher experimental setup of attacks. We evaluate the proposed approache with ResNet18 on multiple benchmark datasets, including CIFAR-5\footnote{A subset of CIFAR-10 that consists of data with label 0 to 4 in CIFAR-10.}, CIFAR-10, CIFAR-100, and Tiny Image Net. 
%我們實驗贏在哪
Experimental results show our proposed approach not only successfully resists all kinds of removal attack, but also has better robustness in all different attack settings compared to other state-of-the-art baselines, especially in distillation attack. 
%After distillation, our approach can still obtain 99.34\% watermark accuracy from the student model while most watermarking methods fail, and can achieve good results even on student models with different architectures. In addition, the watermarking performance stability of our proposed approach under different attacks is better than another method that can also withstand distillation attack, CosWM~\cite{charette2022cosine}, which may suffer a sudden drop in watermarking performance during the fine-tuning attack or during the pruning with re-training attack.
Besides, our methods also have excellent trade-off between the main task and watermarking compared to other baselines and has great functionality-preserving.
%which only has 0.59\% drops on the main task accuracy. Furthermore, the customization ability evaluation demonstrates that our approach has good customization ability and can be generalized to datasets with different numbers of labels.

%我們的貢獻
The contributions are summarized as follows.
\begin{itemize}
  \item We propose a novel deep neural network watermarking framework, \emph{Unified Soft-label Perturbation (USP)}, which not only can resist common watermark removal attacks, but also is robust to distillation attack.
  
  \item We further propose \emph{Customized Soft-label Perturbation (CSP)} to embed customizable watermark by adding specific perturbations to model prediction probability distributions.
  
  \item Experimental results on multiple benchmark datasets show that our approach can be generalized to datasets with different numbers of labels and has good customization ability.
  
  \item Experimental results compared with other baselines demonstrates that our approach has the best trade-off between the main task and watermarking. Besides, even under difficult attack settings, our approach still has the best robustness.
  
  \item Experimental results also indicate our approach has great watermarking performance stability under different attacks. It does not suffer from the sudden drop in signal strength that the signal-based method may encounter.
  
\end{itemize}

\section{Related Works}

Existing deep neural network (DNN) watermarking approaches can be mainly categorized into three types: i) \emph{feature-based methods}, ii) \emph{trigger-set-based methods} and iii) \emph{signal-based methods}.% In this section, we introduce these techniques and  explain why they fail after distillation.

\noindent \textbf{Feature-based.} This type of methods embed designated pattern, such as binary bit string or vector, into the DNN parameters as watermark. Uchida et al.~\cite{uchida2017embedding} embeds watermark by adding a new regularization terms. Fan et al. ~\cite{fan2019rethinking} and Zhang et al.~\cite{zhang2020passport} use \textit{passport-layers} to carry the watermarks. \emph{RIGA}~\cite{wang2021riga} embeds the watermark in the parameters of model, and extract it with a trained DNN.

%However, such methods are very dependent on certain parameters carrying watermark information in the model. Once those parameters are significantly changed or discarded, such as by fine-tuning with large learning rate, heavy pruning or by distillation of knowledge into a model with a completely different architecture, the embedded watermark will be destroyed.

\noindent \textbf{Trigger-set-based.} Trigger-set is a kind of adversarial training samples with specific labels, and the trigger-set-based methods rely on these samples to embed watermarks into DNN. \emph{WNN} proposed by Adi et al.~\cite{adi2018turning} and the embedding framework proposed by Zhang et al.~\cite{zhang2018protecting} both include the trigger-set as watermarks into the training data.

%Just as feature-based methods rely on specific parameters in the DNN model, the trigger-set-based method is very dependent on the trigger set and the knowledge retention ability of DNN model. When a model has good knowledge retention ability, it does not forget old knowledge about the main task while learning new knowledge, the triggering behavior designed in advance. Therefore, the \textit{catastrophic forgetting phenomenon} is a big weakness of this kind of methods. When the model is fine-tuned by the adversaries using its own dataset $D_{a}$ and a larger learning rate, the embedded watermark can be easily removed. Besides, if the adversary using knowledge distillation, only the main task related data would be input, the designated behaviors of trigger-set will never be learned by the malicious model.

\noindent \textbf{Signal-based.} The signal-based methods inject signal into the model output as watermark. Charette, Chu et al.~\cite{charette2022cosine} designed \textit{CosWM} which embeds cosine signal into the output of the watermarked model. However, \textit{CosWM} needs to calculate linear projections, periodic signal function, and power spectrum when embedding and extracting watermark, which is more time-consuming and complicated for users. Moreover, on their paper, only performed \textit{CosWM} on half of the CIFAR-10 dataset and did not experiment with defense against fine-tuning and pruning attacks. %Thus, we implement \textit{CosWM} ourselves with ResNet18 on the whole CIFAR-10 dataset, and apply fine-tuning and pruning attacks to it in our preliminary analysis. According to Table \ref{table:pre2}, its watermark accuracy is not very high on the complete CIFAR-10 dataset, and the effect drops obviously after pruning attack.

In addition to the above methods embedding watermark information into models, there are also some researches designing watermarking mechanisms through APIs, such as \emph{PRADA}~\cite{juuti2019prada}, \emph{Fingerprinting}~\cite{lukas2019deep}, and \emph{DAWN}~\cite{szyller2021dawn}. 
%These methods build on the situation that the trained model is deployed as a service and the users can access it through the API.
%\emph{PRADA}~\cite{juuti2019prada} is designed to analyze query distributions and detect the potential distillation activity. 
%When the query distributions deviate from benign distributions, \emph{PRADA} detects the potential distillation activity.
%\emph{Fingerprinting}~\cite{lukas2019deep} makes backdoor images more transferable through figuring out common adversarial images.% which trigger the same designed prediction on a teacher model and any student model.
%\emph{DAWN}~\cite{szyller2021dawn} is proposed to generate trigger-set from a small subset of querying image by giving the incorrect outputs through API.% Then the model owner could use the trigger-set generated to verify the ownership.
%Although these methods can defend against model stealing attacks similar to knowledge distillation, it is not enough. 
However, API methods is not defensive enough. Since the model itself has no watermark, once someone has a way to obtain the model, it will be directly stolen without any protection.

\section{Problem Definition}

\noindent \textbf{Neural network watermarking.} Watermarking is a method used to protect intellectual property, providing the owner with the ability to claim the ownership. %, and its application in protecting media such as video and images is very common. 
%Extending it to machine learning provides a means of defending against model theft, which does not prevent adversaries from stealing models, but provides the owner with the ability to claim the ownership.
Neural network models are often over-parameterized, and this property can be exploited to embed the information we want other than the main task.
A neural network watermarking scheme basically consists two main steps: i) \textbf{Embedding} and ii) \textbf{Verification}. Let $M$ and $k$ denote the model to be trained and the information to be embed. Then we can have 

\begin{center}
$M_{wm} = Embed(M, k)$,
\end{center}

\noindent where $M_{wm}$ is the embedded model. Let $M_{v} $ denotes the model to be verified, and $V_{wm}$ denotes the return value of the verify step. In the verification step, usually calculate how accurately $k$ can be extracted from the model $M_{v} $ or how well it matches the particular behaviors when $M_{v} $ triggered by the trigger-set. Then this step can be written as

\begin{center}
$V_{wm} = Verify(M_{v}, k)$,
\end{center}

\noindent and the higher the value of $V_{wm}$ is, the better the detection of the presence of watermark information $k$.

%\noindent \textbf{Attacks against neural network watermarking. }
\noindent \textbf{Fine-tuning Attack. }
%Fine-tuning attack is one of the most common watermark attacks considered in existing works.
The adversaries fine-tune the embedded model $M_{wm}$, attempting to erase the embedded watermark information by updating the parameters with their own dataset $D_{a}$. Most of the watermarking methods claim that they are robust to this kind of attack. However, REFIT~\cite{chen2021refit} pointed out that some methods can resist the fine-tuning attack only when the learning rate is small. %The authors experimented with using large learning rates to update the watermarked model~\cite{uchida2017embedding} parameters, and successfully destroyed the embedded watermark.

\noindent \textbf{Pruning Attack. }
%Another common watermark attack is pruning attack. 
Pruning is a post-processing operation of neural network. It removes some connections between neurons to reduce the number of parameters. By pruning, the adversaries may remove some parameters carried watermark information, thus erasing the watermark in $M_{wm}$.

\noindent \textbf{Distillation Attack. } Knowledge distillation~\cite{hinton2015distilling} is one of the model compression methods. 
%It uses the knowledge from trained model, \textit{the teacher model}, to train a new model, \textit{the student model}, which having smaller size. 
For each input, the technique makes the student model imitate the output behavior of the teacher model as closely as possible to learn better than it does on its own. Through distillation, the adversaries can replicate a high-performance new model with their own dataset $D_{a}$ without accessing parameters of $M_{wm}$. Thus, the adversaries will not obtain the parameters carrying the watermark information nor learn any special behavior.

\noindent \textbf{Problem definition. }
Table \ref{table:methodstoattacks} summaries the robustness of different neural network watermarking methods to different attacks from our preliminary analysis in Section 4. From Table \ref{table:methodstoattacks}, we can find that the existing methods cannot resist the above three attacks well at the same time.
Thus, \textbf{our task is to design a robust watermarking method against not only fine-tuning and pruning but knowledge distillation}, making the watermark still exist in the replicated model after attacked by the adversary. Besides, there are some requirements need to met:
\begin{itemize}
\item \textbf{\emph{Functionality-Preserving}}: The impact of watermark embedding on the performance of the main task needs to be as less as possible.

\item \textbf{\emph{Identification Criteria}}: the criteria for success in identifying or extracting watermarks should be clear to prevent disputes. For example, $V_{wm}$ , the return value of the verify step, should be higher than a threshold $V_{thres}$, then we can say the watermark is identified.

\item \textbf{\emph{Robustness against Removal Attacks}}: After the model undergoes watermark removal attacks, the watermark should still be well preserved in the model.

%\item \textbf{\emph{Customization Ability}}: The watermark should be customizable. Different watermarks will not be misidentified to be each other.
\end{itemize}

\begin{center}
\begin{table}[ht]
\caption{Robustness of neural network watermarking methods to different attacks. The mark \cmark $ $ means the watermarking method is robust to the attack, the mark \xmark $ $ means the watermarking method can resist the attack, and the mark $\Delta$ means the watermarking method under this attack is not very good but acceptable, or cannot resist under some attack settings.}
\label{table:methodstoattacks}
\resizebox{\columnwidth}{!}{%
\def\arraystretch{1.6}
\begin{tabular}{ |c||c|c|c|c| } 
\hline
\textbf{Method}& \textbf{Fine-tuning} &  \textbf{Pruning}  &  \textbf{Re-training}  &  \textbf{Distillation}\\
\hline
\texttt{WNN} (USENIX'2018)& $\Delta$ & $\Delta$ & \xmark & \xmark  \\ 
\texttt{DeepIPR} (NIPS'2019)& $\Delta$ & \cmark & \xmark & \xmark  \\ 
\texttt{PA} (NIPS'2020)& $\Delta$ & \cmark & \xmark& \xmark  \\ 
\texttt{RIGA} (WWW'2021)& \cmark & \cmark & \cmark & \xmark  \\ 
\texttt{DAWN} (MM'2021) & \xmark & \xmark & \xmark & \cmark  \\ 
\texttt{CosWM} (AAAI'2022) & \cmark & \xmark & $\Delta$ & \cmark  \\ 
%\textbf{Ours} & \cmark & \cmark & \cmark  \\
\hline
\end{tabular}%
}
\end{table}
\end{center}

\section{Preliminary Analysis}
\label{pre}

\noindent In this section, we conduct preliminary analyses with recent watermarking approaches to know their performance and robustness under different settings of removal attacks.
%including \emph{Fine-tuning Attack}, \emph{Pruning Attack}, and \emph{Distillation Attack}.
For each approach, we generate a watermarked ResNet18 model on CIFAR-10 with their method respectively. Each watermarked model is trained with 160 training epochs and has a learning rate starting as 0.1.% and decreased by 0.1 times when 80 epoch and 120 epoch.

\noindent \textbf{Functionality-Preserving.} Table \ref{table:pre1} shows the ability of functionality-preserving of each watermarking method. It can be found that most methods achieve good results on main task accuracy and watermark accuracy. The trigger-set-based method, \texttt{WNN}, has relatively poor functional-preserving. 
%compared to the other two based methods. 
Because trigger-set-based methods would combine trigger set into the main task training set, data not relevant to the main task may affect the performance of the model in the main task. 
Besides, the overall performance of \texttt{DAWN} is significantly lower. The reason should be that the way \texttt{DAWN} generates trigger-set is to directly modify the labels of some data in the main task dataset, so that the features of trigger-set will be similar to the features of main task dataset but having different labels. %Therefore, the performances of main task and trigger-set will affect each other, so that the effect of neither will be high. 

\begin{table}[ht]
\centering
\caption{Functionality-preserving of different neural network watermarking methods on CIFAR-10}
\label{table:pre1}
\resizebox{\columnwidth}{!}{%
\def\arraystretch{1.6}
\begin{tabular}{c|c|cc}
\multicolumn{1}{c}{\textbf{}} & \multicolumn{1}{c}{\textbf{Clean}} & \multicolumn{2}{c}{\textbf{Watermarked}}\\
\multicolumn{1}{c}{\textbf{Method}} & \multicolumn{1}{c}{main acc.(\%)} & \multicolumn{1}{c}{main acc.(\%)} & \multicolumn{1}{c}{wm acc.(\%)}\\
\hline

\texttt{WNN} (USENIX'2018)& & \textit{\textbf{90.67 (-4.63)}}& 100.00\\

\texttt{DeepIPR} (NIPS'2019)& & \textit{\textbf{92.11 (-3.19)}}& 100.00\\

\texttt{PA} (NIPS'2020)& \multirow{2}{*}{95.30} & 94.02 (-1.27)& 100.00\\

\texttt{RIGA} (WWW'2021)& ~ & 94.43 (-0.87)& 100.00\\

\texttt{DAWN} (MM'2021)& &\textit{\textbf{79.27 (-26.1)}} & 88.80\\

\texttt{CosWM} (AAAI'2022)& & 95.14 & {88.28}\\

\end{tabular}%
}
\end{table}

\noindent \textbf{Fine-tuning Attack.} Table \ref{table:pre2} shows the robustness to fine-tuning attacks with different learning rate settings of each watermarking method. We fine-tune the watermarked models for 100 epochs with small learning rate, $0.001$ and larger learning rate, $0.01$ on the same CIFAR-10 training set. 
%Under the fine-tuning attack with small learning rate, the watermark of most methods can still be maintained very well. 
The watermark accuracy of \texttt{DAWN} is decreased since the same reason mentioned above: the trigger-set is generated from the main task dataset. Thus, after fine-tuning, the effect caused by the trigger-set on the main accuracy is decreased, and watermark accuracy drops dramatically.
Then we look at the results under the attack with a larger learning rate. It can be found that the watermark embedded in almost all methods are obviously damaged.% Especially trigger-set based method, when its watermark is destroyed, the accuracy of the main task does not decrease but increases.

\noindent \textbf{Pruning Attack.} Table \ref{table:pre2} shows the robustness to pruning attacks with and without re-training of each watermarking method. we adopt the classic pruning method~\cite{han2015deep} proposed and show the results after pruning off 80\% parameters.% Since under this setting, the watermarking accuracy of most watermarking methods drops with the accuracy of the main task at the same time. 
However, after re-training 100 epochs with a learning rate of 0.01, we can find the watermark accuracy of most methods drop obviously but their main task accuracy recovers with re-training. %This results are reasonable, because the effect of re-training is very similar to fine-tuning. If the watermarking method cannot resist fine-tuning, the embedded watermark will also be destroyed after re-training.

\noindent \textbf{Distillation Attack.} Table \ref{table:pre2} shows the robustness in distillation attack of each watermarking method. We distill the watermarked model with the same CIFAR-10 dataset, and set ResNet18 as student model, training epochs as 160, learning rate as 0.1, and distillation temperature as 4. 
%First, the watermarks of the two methods \texttt{DeepIPR} and \texttt{PA} disappear completely, since the watermark is embedded on the passport-layer added by themselves. After distilling the watermarked model to a student model of other architectures, there will be no way to find where the watermark is. 
%Then, w
We can see that except for the signal-based method, \texttt{CosWM}, and API method, \texttt{DAWN}, the watermarks of all the other methods cannot be well preserved on the distilled student model. Because knowledge distillation attack can replicate the model with just the main task related dataset and without accessing parameters of the watermarked model. Thus, the student model will not obtain the parameters carrying the watermark information embedded by featured based methods nor learn any special behavior of trigger-set-based methods.

\begin{table*}[ht]
\centering
\caption{Robustness to removal attacks of different watermarking methods on CIFAR-10. The main (\%) indicates main task accuracy and the wm (\%) indicates watermark accuracy.}
\label{table:pre2}
\begin{adjustbox}{width=\textwidth}
\def\arraystretch{1.6}
\begin{tabular}{cccccccccccccc}
%\toprule
\multicolumn{1}{c}{ } & \multicolumn{2}{c}{\textbf{Before Attack}} & \multicolumn{2}{c}{\textbf{Fine-tune (lr = 0.001)}} & \multicolumn{2}{c}{\textbf{Fine-tune (lr = 0.01)}}& \multicolumn{2}{c}{\textbf{Prune}} & \multicolumn{2}{c}{\textbf{Re-train after Prune}} & \multicolumn{2}{c}{\textbf{Distill}}\\
\cmidrule(rl){2-3} \cmidrule(rl){4-5} \cmidrule(rl){6-7} \cmidrule(rl){8-9} \cmidrule(rl){10-11} \cmidrule(rl){12-13}
\textbf{Method} & {main (\%)} & {wm (\%)} & {main (\%)} & {wm (\%)} & {main (\%)} & {wm (\%)} & {main  (\%)} & {wm (\%)} & {main (\%)} & {wm (\%)} & {main (\%)} & {wm (\%)} \\
%\hline
\midrule
\texttt{WNN} (USENIX'18)& 90.67 & 100.00 & 90.72 & 100.00 & 92.43 & \underline{\textit{\textbf{52.31}}}& 88.59 & 75.61 & 88.38 & \underline{\textit{\textbf{38.00}}}& 92.24 & \underline{\textit{\textbf{8.00}}}\\
\texttt{DeepIPR} (NIPS'19)& 92.11 & 100.00 & 92.40 & 100.00 & 92.55 & \underline{\textit{\textbf{70.48}}}& 88.27 & 94.87 & 94.34 & \underline{\textit{\textbf{69.53}}}& 90.14 & \underline{\textit{\textbf{0.00}}}\\
\texttt{PA} (NIPS'20)& 94.02 & 100.00 & 94.07 & 100.00 & 92.26 &\underline{\textit{\textbf{70.73}}}& 55.04 & 96.21 & 91.33 &\underline{\textit{\textbf{69.94}}}& 92.33 & \underline{\textit{\textbf{0.00}}}\\
\texttt{RIGA} (WWW'21)& 94.43 & 100.00 & 94.97 & 100.00 & 93.34 & 99.24 & 77.98 & 100.00  & 93.40 & 100.00 & 94.12  & \underline{\textit{\textbf{0.00}}}\\
\texttt{DAWN} (MM'21) & 79.27 & 88.80 & 80.53 & \underline{\textit{\textbf{8.80}}} & 80.34 & \underline{\textit{\textbf{0.00}}}& 11.50 & \underline{\textit{\textbf{12.80}}} & 78.48 & \underline{\textit{\textbf{0.00}}} & 78.11 & 86.20\\
\texttt{CosWM} (AAAI'22)& 95.14 & 88.28 & 94.95 & 88.35 & 95.00 & 82.97& 52.45 & \underline{\textit{\textbf{57.39}}} & 94.59 & \underline{\textit{\textbf{75.85}}}& 95.45 & 83.54\\

%\bottomrule

\end{tabular}
\end{adjustbox}
\end{table*}

\noindent From our preliminary analysis, we can observe two things: 
\begin{itemize}
    \item First of all, \textbf{most of the existing watermarking methods can not resist to tougher removal attack setting}, such as fine-tuning with larger learning rate and re-training after pruning.
    \item Secondary, \textbf{most of the watermarks embedded by existing watermarking methods will disappear after being attacked by knowledge distillation}.
\end{itemize}

\section{Algorithm Design}
\label{sec:algo}

Most of the existing watermarking methods can not defend against knowledge distillation attack. 
%We believe that the main reason is that the adversary can replicate the trained model by imitating the prediction probability distribution of the victim without access parameters. 
In order to address this problem, we design a watermarking framework to hide watermarks in the prediction probabilities to resist the imitation behavior in knowledge distillation.

\subsection{Unified Soft-label Perturbation (USP)} %\shenr{Unified Soft-label Perturbation (USP)}

The basic idea of \emph{USP} is to add some perturbation as a watermark to the output of the model during the training process, making the prediction probability distribution of the trained model different from general models. In addition, we design a detector model to detect the differences in the outputs to identify the watermarked model. If someone intends to steal the trained model by knowledge distillation, the perturbation effects will also be transferred to the model replicated, because the student model would learn the output behavior of the teacher model. %In this way, the owner of the original model can check suspect model by detecting its output.

\begin{figure}[ht]
\centering
\includegraphics[width=1.0\columnwidth]{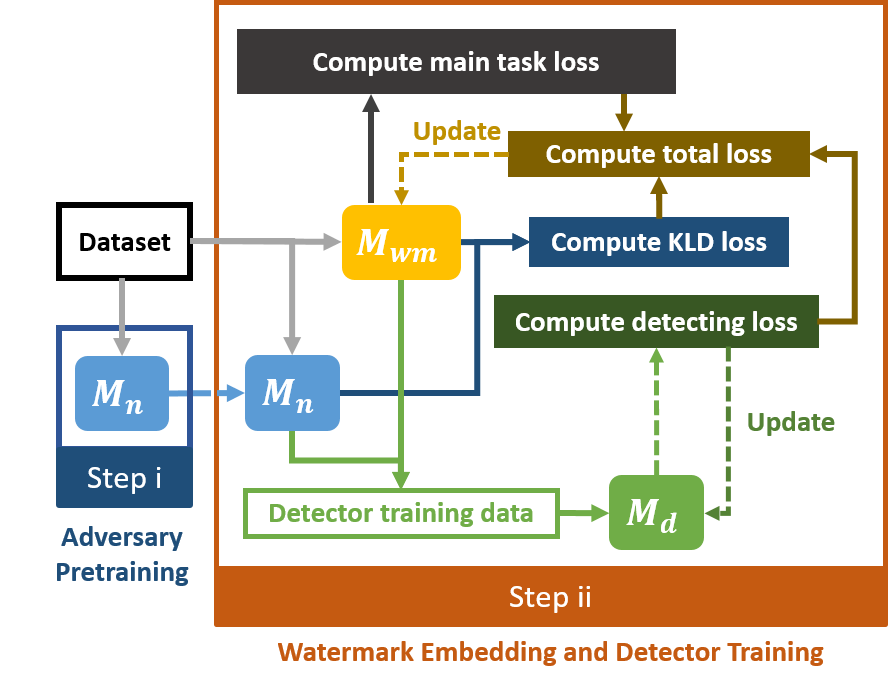} % Reduce the figure size so that it is slightly narrower than the column. Don't use precise values for figure width.This setup will avoid overfull boxes.
\caption{The flow of \emph{Unified Soft-label Perturbation (USP)}.}

\label{fig:basic_flow}
\end{figure}

The flow is shown in Figure \ref{fig:basic_flow}. There are two main steps: \textbf{i)\textit{ adversary pretraining}} and \textbf{ii) \textit{watermark embedding and detector training}}. First, in step i, without adding any perturbation, we train a normal model $M_{n}$ to be an adversary in the next step. The role of the pre-trained model $M_{n}$ in step ii is to help train the detector to distinguish the difference between the watermarked model and the normal model.

In step ii, we designed a training method to train the model $M_{wm}$ to be watermarked and the corresponding detector $M_{d}$. To train a model having special prediction probability distribution, we apply the KL divergence loss function with soft temperature $\mathcal{T}$, similar to knowledge distillation, to calculate the difference between the outputs of it and the adversary model $M_{n}$, the normal model trained in step i. 
\begin{equation}\label{algo:eq-kl}
\mathcal{L}_{KLD}= \mathcal{T}^{2} \cdot KL( log\_softmax( O_{n}/\mathcal{T}) , softmax(O_{wm}/\mathcal{T}) ),
\end{equation}
where $O_{wm}$ and $O_{n}$ are respectively the output of $M_{wm}$ and the output of $M_{n}$. The purpose of using a loss function similar to knowledge distillation is to hope that when an adversary tries to distill the model, the perturbations in the model output can be more easily learned by the student model.

\begin{figure*}[t]
\centering
\includegraphics*[width=\textwidth]{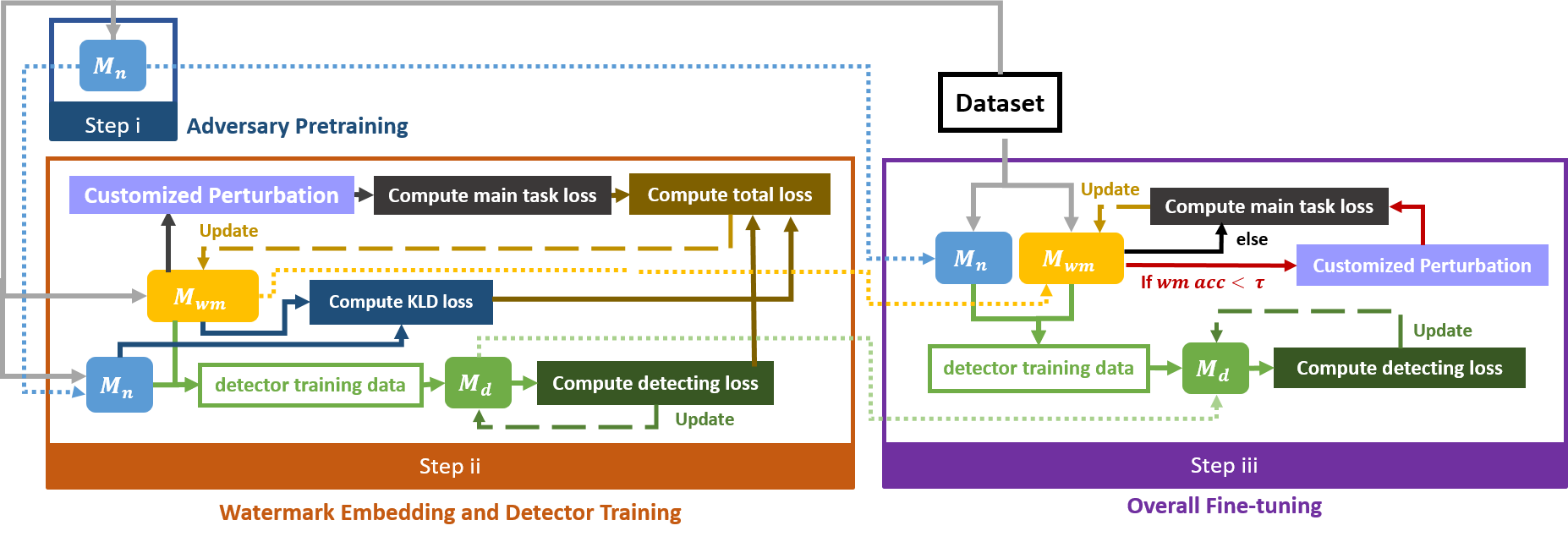} % Reduce the figure size so that it is slightly narrower than the column. Don't use precise values for figure width.This setup will avoid overfull boxes.
\caption{The flow of \emph{Customized Soft-label Perturbation (CSP)}.}

\label{fig:meth_MAP}
\end{figure*}

Afterwards, we the add the KL divergence loss with a negative coefficient $\alpha$ to the main task loss calculated by cross-entropy to get the total model loss $\mathcal{L}_{Model}$ which is defined as follows.
\begin{equation}\label{algo:eq-model-basic}
\mathcal{L}_{Model}^{USP}= \mathcal{L}_{MainTask}^{USP} + \alpha \cdot \mathcal{L}_{KLD}.
\end{equation}
\begin{equation}\label{algo:eq-main-basic}
\mathcal{L}_{MainTask}^{USP}= CrossEntropy( O_{wm} ,Y),
\end{equation}
where $Y$ is the ground-truth label. The negative coefficient $\alpha$ allows model $M_{wm}$ to increase the difference from the adversary $M_{normal}$ as much as possible, thus promoting it to have different prediction probability distributions. 

The detector model is also trained when training the model. We design a five-layer MLP as the detector, and take the output of $M_{wm}$ and $M_{n}$ as its input to perform the binary classification task of identifying watermarks. In each epoch, the main task data $x^{i}$ $\in$ $X$ will be input into $M_{wm}$ and $M_{n}$ at the same time, generating the prediction outputs $o_{wm}^{i}$ and $o_{n}^{i}$. For each $o_{wm}^{i}$, we assign label $l_{wm}^{i} = 1$, likewise, for each $o_{n}^{i}$, we assign label $l_{n}^{i} = 0$.
\begin{equation}
\label{algo:assign-label}
\textbf{Label} (o_{m}^{i}) = \Big \{
	  \begin{tabular}{cc}
	   $l_{wm}^{i}$ = 1 & if $m$ is $M_{wm}$,\\
	   $l_{n}^{i}$ = 0 & if $m$ is $M_{n}$.
	  \end{tabular}
	 \\
\end{equation}

Then we can have detector training dataset $D_{d}$ = ($X_{d}$, $Y_{d}$), where $X_{d} = \left\{ O_{wm}, O_{n}\right\}$ is the input and $Y_{d} = \left\{ L_{wm}^{i}, L_{n}^{i}\right\}$ is the label. After preparing the dataset, the binary classification training of the detector can be performed. For this task, we using binary cross-entropy to calculate the detecting loss $\mathcal{L}_{Detect}$. Besides, to reduce additional computational cost, we set an early stopping mechanism for the detector training process, which ends the detector training once the detecting accuracy reaches 100.
\begin{equation}\label{algo:eq-detect}
\mathcal{L}_{Detect}= BinaryCrossEntropy( M_{d}(X_{d}) ,Y_{d}),
\end{equation}

After updating the detector model $M_{d}$ with $\mathcal{L}_{Detect}$, we add $\mathcal{L}_{Detect}$ to the model loss $\mathcal{L}_{Model}$ to get a total loss $\mathcal{L}_{Total}$ for updating $M_{wm}$. In this way, $M_{wm}$ will also take into account the discriminative ability of the detector when updating. We define $\mathcal{L}_{Total}$ as follows.

\begin{equation}\label{algo:eq-total-basic}
\mathcal{L}_{Total}^{USP}= \mathcal{L}_{Model}^{USP} + \beta \cdot \mathcal{L}_{Detect},
\end{equation}
where $\beta$ is the coefficient to moderate the influence of the detector. Then, we finally obtain the watermarked model trained by \emph{USP}. When the model owner train a self-used model, \emph{USP} can protect the IP of the model well from not only fine-tuning attack and pruning attack, but distillation attack, and can also be used to detect whether the model has been maliciously changed since the watermark in the model is related to performance the model.

\begin{center}
\begin{table}[ht]
\caption{Result of three different watermarked model and detector pairs generated by \emph{USP} on CIFAR-10.}

\label{table:basic_method}
\resizebox{\columnwidth}{!}{%
\def\arraystretch{1.6}
\begin{tabular}{ c|c|c|c } 

 & $M_{wm1}$ &  $M_{wm2}$ & $M_{wm3}$\\ \hline\hline

Main task acc. of Unwatermarked Model (\%) & \multicolumn{3}{c}{ 95.30 } \\
\hline
Main task acc. of $M_{wm}$ (\%) & 94.10 & 93.00 & 93.46 \\ 
\hline
Identified by $M_{d1}$ (\%) & \textbf{99.74} & 93.91 & 87.81 \\ 
Identified by $M_{d2}$ (\%) & 99.87 & \textbf{97.56} & 91.14 \\ 
Identified by $M_{d3}$ (\%) & 99.91 & 98.88 & \textbf{99.57}
\end{tabular}%
}
\end{table}
\end{center}

However, if there is a need for customizing, \emph{USP} is defective: the detector is \textbf{not sensitive enough to identify the custom characteristics of different watermarks}. There are the results of three different watermarked model and detector pairs generated by the \emph{USP} method in Table \ref{table:basic_method}  From Table \ref{table:basic_method}, we can find although the detector can well detect whether a model is watermarked, it can not correctly distinguish whether the model has someone's watermark. \textbf{We think the reason is that only using KL divergence loss can not constrain the form of the perturbation well}, which may cause the model just randomly changes the probability distribution to reduce the total loss, \textbf{making the characteristics of the output distribution inconsistent}. Thus, the detector can only distinguish the difference from the normal model, instead of the specific watermark features. 
%Besides, the \textbf{functionality-preserving has room for improvement}, we can find from Table \ref{table:basic_method} that each watermarking model has at least a 1.2\% reduction in main task accuracy. 
To improve the deficiency, we design \textbf{\emph{Customized Soft-label Perturbation (CSP) }}.

\subsection{Customized Soft-label Perturbation (CSP)} %\shenr{改為Costomized Soft-label Perturbation (CSP)}
\label{sec:algo_MAP}
When we train a model and are going to authorize others to use the model, we can use \emph{CSP} to protect IP. We customize the watermark for each user, and if there is an authorized model breached by a user, we can use \emph{CSP} to verify the watermark, prove our ownership and find the person who breached the model. 

The flow is shown in Figure \ref{fig:meth_MAP} which having three steps: \textbf{i)\textit{ adversary pretraining}}, \textbf{ii) \textit{watermark embedding and detector training}}, and \textbf{iii) \textit{overall fine-tuning}}. In the first step, CSP also trains an ordinary model $M_{n}$ to become the adversary in the next step. While in the watermark embedding and detector training step, \emph{CSP} has some adjustment. 
We first introduce some terms. We define a vector $S \in {\left\{ 0, 1, -1 \right\}}^{l}$ as the watermark signal, where $l$ is the length of $S$, and $\gamma$ as a signal strength factor. In order to solve the problem of insufficient customization ability of the \emph{USP} method, we try to make the output probability distribution of $M_{wm}$ have consistent features. We adjust and constrain $O_{wm}$ to be $\widetilde{O}_{wm}$ with $S$ and $\gamma$ as follows.
\begin{equation}\label{algo:eq-out-personal}
\tilde{o}_{wm}^{i} = o_{wm}^{i} - \gamma \cdot S,
\end{equation}
where $O_{wm} = \left \{ o_{wm}^{0} \cdots o_{wm}^{n} \right\}$ and $n$ is the number of training data. We preset the length of the watermark, $l$, to be the same as the number of classes of the main task dataset. For example, when using CIFAR-10 as the main task dataset, the length of the $S$ will be set to 10. The factor $\gamma$ is used to moderate the signal strength literally. The strength of the signal will increase as the factor is increased. 

However, when embedding watermarks on a task with many labels, if the length of the $S$ is the same as the number of labels, it will be more complicated to assign. In this case, just filter out part of the labels to carry the watermark, then the length of $S$ can be reduced. For example, on CIFAR-100 task, we just using label 0 to label 9 for embedding as follows.

\begin{equation}\label{algo:eq-filter}
\tilde{o}_{wm}^{i}[j] = \Big \{
	  \begin{tabular}{cc}
	   $o_{wm}^{i}[j]$ - $\gamma$ $\cdot$ S[k] & if $dict[j]$ = $k$,\\
	   $o_{wm}^{i}[j]$ & else,
	  \end{tabular}
	 \\
\end{equation}
where $dict$ is the dictionary to filter the labels, and $dict[j]$ = $k$ means label $j$ is the $k$-$th$ one be filtered out to embed watermark.

Basically, the composition of the model loss function $\mathcal{L}_{Model}$ is as same as that of the \emph{USP} framework in Eq.\ref{algo:eq-model-basic}. We just need to modify Eq.\ref{algo:eq-main-basic} slightly to obtain a new main task loss function for this framework.

\begin{equation}\label{algo:eq-model-per}
\mathcal{L}_{Model}^{CSP}= \mathcal{L}_{MainTask}^{CSP} + \alpha \cdot \mathcal{L}_{KLD}.
\end{equation}

\begin{equation}\label{algo:eq-main-per}
\mathcal{L}_{MainTask}^{CSP}= CrossEntropy( \widetilde{O}_{wm} ,Y)
\end{equation}
We replace $O_{wm}$ with $\widetilde{O}_{wm}$, thus when $M_{wm}$ is to be updated, it will find that the prediction probabilities of certain classes are low, and then tend to increase the prediction probabilities of those classes so that we can achieve the effect of adding fixed perturbations.

For the detector training dataset, we assign labels in the same way as the \emph{USP} method in Eq.\ref{algo:assign-label}, but modify how the detector training data be generated. Here, we use $log\_softmax$ to modify $O_{wm}$ and $O_{n}$ to $\hat{O}_{wm}$ and $\hat{O}_{n}$ .

\begin{equation}\label{algo:eq-dxwm-personal}
\hat{O}_{wm} = log\_softmax( O_{wm} )
\end{equation}
\begin{equation}\label{algo:eq-dxn-personal}
\hat{O}_{n} = log\_softmax( O_{n} )
\end{equation}

We hope the detector can learn the features of the watermark signal we embed according to the bias on the probability distribution. Therefore, we adopt the $log\_softmax$ function that takes the $log$ of the result calculated by $softmax$. $softmax$ can convert the output into the form of probability we want, and taking the $log$ value of the result makes the gap between each probability more obvious.% so that the detector can more clearly identify the features.

The detector training process is as same as the one in the \emph{USP}. After updating the detector model $M_{d}$, we sum the detecting loss $\mathcal{L}_{Detect}$ and the model loss $\mathcal{L}_{Model}^{CSP}$ to obtain the total loss. 
The total loss $\mathcal{L}_{Total}^{CSP}$ can be defined as follow. 
\begin{equation}\label{algo:eq-total-per}
\mathcal{L}_{Total}^{CSP}= \mathcal{L}_{Model}^{CSP} + \beta \cdot \mathcal{L}_{Detect}
\end{equation}

In overall fine-tuning step, we fine-tune the watermarking model $M_{wm}$ and continue to train the detector while fine-tuning in this step. We regard fine-tuning as a data augmentation of detector. Here, we updated $M_{wm}$ without adding KL loss, and use the original output ${O}_{wm}$ without modification. The loss function of fine-tuning $\mathcal{L}^{Finetune}$ is defined as follow.

\begin{equation}\label{algo:eq-finetune}
\mathcal{L}^{Finetune} = CrossEntropy( {O}_{wm} ,Y)
\end{equation}

The reason we directly fine-tune $M_{wm}$ without modifying its output is that in the last stage of training, $M_{wm}$ will come to a relatively optimal area in the loss function considering signal embedding.%, and will also use a relatively small learning rate. 
We believe that using the original output directly in such cases will allow $M_{wm}$ to focus more on improving the performance of the main task without causing watermark damage.
Moreover, %training the detector while fine-tuning can improve the robustness of the detector. F
fine-tuning will make the prediction distribution have slight changes that the detector has not seen before, so that the detector then can obtain more training materials to enhance its robustness.

However, in order to prevent the situation where fine-tuning has a significant impact on the watermark, we set up a re-embed mechanism. When the watermark accuracy is lower than the threshold $\tau$, $O_{wm}$ will be modified again to deepen the watermark. At last, we summarize the loss function of the overall fine-tuning $\mathcal{L}^{Finetune}$ step as follows:

\begin{equation}\label{algo:overall}
{L}^{Finetune} = \Big \{
	  \begin{tabular}{cc}
	   CrossEntropy( ${O}_{wm}$ ,Y) & if $acc_{wm}$ $>$ $\tau$ ,\\
	   CrossEntropy( $\widetilde{O}_{wm}$ ,Y) & else.
	  \end{tabular}
	\\
\end{equation}

%\subsection{Preliminaries} 
%\label{sec:algo_preliminaries}

% ---
\section{Experimental Results}
\label{sec:experiment}
 
 In this section, we evaluate the proposed methods, including \textbf{Unified Soft-label Perturbation (USP)} and \textbf{Customized Soft-label Perturbation (CSP) }, on four different datasets. We also compare them with classic and state-of-the-art methods: \texttt{WNN, DeepIPR, PA, RIGA, DAWN,} and \texttt{CosWM}. We also evaluate all the approaches on their \textit{functionality-preserving ability} and their robustness under different attack settings.

%,  random horizontal flip, random crop and normalization for training images, and only use normalization for testing images.

\subsection {Baselines.} 
In order to demonstrate the effectiveness and superiority of our approach, we compare our approaches with six methods.

\noindent \textbf{Trigger-set-based}: \texttt{WNN} \textit{(USENIX'18)} uses backdooring method to train a watermarked model with trigger-set. In our experiments, we use the same trigger set as the one released on the github.

\noindent \textbf{Feature-based}: \texttt{DeepIPR} \textit{(NIPS'19)} inserted \textit{passport-layers} they proposed to carry the watermark bit-strings. We implement \texttt{DeepIPR} with the same passport setting as the one released on the github.

\noindent \textbf{Feature-based}: \texttt{PA} \textit{(NIPS'20)} improves the limitation of \texttt{DeepIPR}. Here we also use the same passport setting as the one released on the github.

\noindent \textbf{Feature-based}: \texttt{RIGA} \textit{(WWW'21)} embedded the watermark in the parameters of model, and extract the watermark from model parameters by a trained DNN. Because the source code does not provide ready-made experimental results or settings of ResNet18 on CIFAR-10, we implement \texttt{RIGA} in Pytorch ourselves according to its Tensorflow based source code, and apply it to ResNet18 on CIFAR-10.

\noindent \textbf{Signal-based}: \texttt{CosWM} \textit{(AAAI'22)} is a signal embedding method that embeds cosine signal into the output of model as watermark. However, the source code of \texttt{CosWM} and the complete experiment setting are not provided. We implement it by ourselves and reproduce it as similarly as possible. In our experiments, we select label $0$ as target label, set the angular frequency as $10.0$, set the the signal amplitude  as $0.2$, and random generate a projection vector. 

\noindent \textbf{API mechanisms}: \texttt{DAWN} \textit{(MM'21)} is a watermarking mechanisms performing on the API platform. Since the source code provided on the github does not offer ready-made experimental results or settings of ResNet18 on CIFAR-10, we use the source code and slightly modify the settings of ResNet34 on CIFAR-10 they provide, change the model architecture from ResNet34 to ResNet18.

\begin{table*}[ht]
\centering
\caption{Performance Evaluation Overview on Different Neural Network Watermarking Methods on CIFAR-10. The main (\%) indicates main task accuracy and the wm (\%) indicates watermark accuracy.}
\label{table:exp_overview}
\begin{adjustbox}{width=\textwidth}
\def\arraystretch{1.6}
\begin{tabular}{ccccccccccc}
\toprule
\multicolumn{1}{c}{\textbf{Clean model acc. (\%):} 95.30} & \multicolumn{2}{c}{\textbf{Original}} & \multicolumn{2}{c}{\textbf{Fine-tune (lr=0.01)}} & \multicolumn{2}{c}{\textbf{Prune}} & \multicolumn{2}{c}{\textbf{Re-train after prune}} & \multicolumn{2}{c}{\textbf{Distill}} \\
\cmidrule(rl){2-3} \cmidrule(rl){4-5} \cmidrule(rl){6-7}  \cmidrule(rl){8-9} \cmidrule(rl){10-11} 
 \textbf{Method} & {main acc (\%)} & {wm acc (\%)} & {main acc (\%)} & {wm acc (\%)} & {main acc (\%)} & {wm acc (\%)} & {main acc (\%)} & {wm acc (\%)} & {main acc (\%)} & {wm acc (\%)} \\
\hline
\midrule
 \texttt{WNN} (USENIX'18) & 90.67 (\underline{\textit{-4.63}}) & 100.00 & 92.43 & \underline{\textit{52.31}} & 88.59 & 80.00 & 88.38 & \underline{\textit{38.00}} & 92.24 & \underline{\textit{8.00}}\\
 
 \texttt{DeepIPR} (NIPS'19)& 92.11 (\underline{\textit{-3.19}}) & 100.00 & 92.55 & \underline{\textit{70.48}} & 88.24 & 94.87 & 95.34 & \underline{\textit{69.53}} & 90.14 & \underline{\textit{0.00}}\\
 
 \texttt{PA} (NIPS'20)& 94.02 (-1.27) & 100.00 & 92.26 & \underline{\textit{70.73}} & 55.04 & 96.21 & 91.33 & \underline{\textit{69.94}} & 92.33 & \underline{\textit{0.00}}\\ 
 
 \texttt{RIGA} (WWW'21)& 94.43 (-0.87) & 100.00 & 93.34 & 100.00 & 77.98 & 100.00 & 93.40 & 100.00 & 94.12 & \underline{\textit{0.00}}\\
 
 \texttt{DAWN} (MM'21)& 79.27 (\underline{\textit{-26.1}}) & \underline{\textit{88.80}} & 80.34 & \underline{\textit{0.00}} & 11.50 & \underline{\textit{12.80}} & 78.48 & \underline{\textit{0.00}} & 78.11 & 86.20\\
 
 \texttt{CosWM} (AAAI'22)& 95.14 (-0.16) & \underline{\textit{88.28}} & 95.00 & 82.97 & 52.45 & \underline{\textit{57.39}} & 94.59 & \underline{\textit{75.85}} & 95.15 & 83.54\\
 
 \hline
 \textbf{USP (Ours)}& \textbf{93.75 (-1.55)} & \textbf{99.56} & 93.84 & \textbf{86.28} & 51.49 & \textbf{99.37} & 94.10 & \textbf{77.12} & 94.92 & \textbf{99.27}\\
 
 \textbf{CSP (Ours)}& \textbf{94.71 (-0.59)} & \textbf{98.68} & 94.61 & \textbf{88.72} & 34.46 & \textbf{90.80} & 94.19 & \textbf{87.40} & 90.05 & \textbf{91.86}\\
 
\bottomrule 
\end{tabular}
\end{adjustbox}
\end{table*}

\subsection{Experiment Setting.}
\noindent \textbf{Baselines.} For fair comparisons, we train a ResNet18 on CIFAR-10 with the same training epochs, starting learning rate, and batch size for each baseline method as possible. We set training epochs as $160$ and starting learning rate as $0.1$.

\noindent \textbf{USP.} We train ResNet18 on CIFAR-10 with $160$ training epochs,  $0.1$ starting learning rate.

\noindent \textbf{CSP.} We train ResNet18 on four datasets. On CIFAR-5 and CIFAR-10, we set training epochs as $120$ and fine-tuning epochs as $120$. The learning rate starts from $0.1$ and decreases by timing $0.1$ in epoch $80$, $120$. On CIFAR-100 and Tiny ImageNet, we set training epochs as $150$ and fine-tuning epochs as $120$. The learning rate of CIFAR-100 task starts from $0.1$ and decreases by timing $0.2$ in epoch $60$, $120$ and $150$. The learning rate of Tiny ImageNet task starts from $0.001$ and also decreases by timing $0.2$ in epoch $60$, $120$ and $160$.

\noindent The architecture of detector is a 5-layer MLP. The detector optimizer is Adam, and the starting learning rate is $0.008$. Besides, only when the watermark accuracy is greater than 85\%, we consider the watermark detection successful.

\subsection {Terms and performance metrics.}

We evaluate the approaches with the following metrics. 

\noindent \textbf{\emph{main acc.}} indicates the accuracy on the main task.

\noindent \textbf{\emph{wm acc.}} For our method, \emph{wm acc.} indicates the accuracy of the detector detecting the outputs from the watermarked and unwatermarked model. The higher the value indicates the better the detector can identify whether there is a watermark in the outputs. If there is $x$ testing image, we will have $2x$ balanced detection testing data, of which there are $x$ watermarked data and $x$ unwatermarked data each. For example, CIFAR-10 has $10,000$ testing images, we will obtain $20,000$ testing detection testing data, of which there are $10,000$ watermarked data and $10,000$ unwatermarked data. 
%\shenr{講一下wm acc.怎麼算，以cifar 10舉例，說testing image有x張我們就做x次detection。然後要說這個data是balanced的}

For \texttt{WNN} and \texttt{DAWN}, \emph{wm acc.} indicates the accuracy of the trigger-set. The higher the value, the better the classification result of the trigger set, that is, the better the watermark identification effect.

For \texttt{DeepIPR}, \texttt{PA}, and \texttt{RIGA}, \emph{wm acc.} indicates the bit-correct-rate of the embedded signature bit-strings. A higher value indicates the signature can be identified better.

For \texttt{CosWM}, \emph{wm acc.} indicates the signal-noise-ratio of the embedded signal. A higher value indicates stronger signal.

\noindent \textbf{\emph{wm det rate.}} indicates the rate of watermark detected by the detector in the outputs of a model. The higher \emph{wm det rate.}, the higher confidence of the detector in recognizing the the model.

\noindent \textbf{\emph{cmps ratio.}} indicates compression ratio, the ratio of the number of parameters of the student model and the teacher model, i.e., the ratio of the student model size to the teacher model size. %A larger compression ratio means that the teacher model is compressed into a smaller student model.

\subsection{Overview}

% 1.我們function preserveing很好
% 2. 我們在大家都壞掉的設定下很好
We compare the robustness of our proposed approaches with the baselines in Table \ref{table:exp_overview}. Here we obtain the watermarked ResNet18 models on CIFAR-10 with each watermarking method, and apply removal attacks with tougher settings, which make most watermarking methods less effective in the preliminary analysis experiments, to them.

\subsubsection{Functionality Preservation.} First of all, we evaluate the functionality-preserving ability of each method. From the experimental results in Table \ref{table:exp_overview}, our proposed approaches have great functionality-preserving ability which only have 1.55\% and 0.59\% drops respectively on the main task accuracy. Compared to other baselines, our proposed approaches have outstanding trade-off between the main task and watermarking.

\subsubsection{Fine-tuning attack.} For fine-tuning attack, we adopt larger learning rate, 0.01, to fine-tune the watermarked models for 100 epochs. We take the result with the highest main acc. From the experimental results, most methods have obviously drop in watermark accuracy, but maintain well or even improve in main task accuracy. In contrast, the watermark accuracy of ours and \texttt{CosWM} is only a little affected. %Besides, our methods have the best robustness that still retain and 86.28\% and 88.72\% watermark accuracy after fine-tuning.

\subsubsection{Pruning attack without re-training.} For pruning experiment, we prune 80 \% weights off without re-training, and the results are shown in Table \ref{table:exp_overview}. From the results, \texttt{Dawn} and \texttt{CosWM} have significant drop on both the main task accuracy and watermark accuracy after pruning, while our methods can still maintain high watermarking accuracy. It can be seen that in pruning attack without re-training, our methods have better robustness.

\subsubsection{Pruning attack with re-training.} %Then, we apply pruning attack with re-training to all the methods and show the results in Table 8.
In this experiment, we also prune 80\% weights off and set the re-training epochs as 100 and the learning rate as 0.01. We take the result with the highest main acc. Though the watermarking accuracy of \texttt{WNN}, \texttt{Deep-IPR}, and \texttt{PA} only have little drop after pruning, their watermarking effect becomes poor after re-training. Moreover, while both the main task accuracy and watermark accuracy of \texttt{CosWM} recover after re-training, the watermark accuracy is much lower then it used to be. Although our methods have a little drop in watermark accuracy, they still maintain high identification ability after re-training. %This indicates that we have better protection than most baselines against pruning attack no matter with or without retraining.

\subsubsection{Distillation attack} %At last, we apply distillation attack. 
We distill the watermarked model to a ResNet18 student model on CIFAR-10 for 160 epoch and set the learning rate as 0.1 and the distillation temperature as 4. We have known that all baselines except signal-based methods are incapable of resisting distillation attacks from the preliminary analysis. %, because by distillation, the adversary can replicate a mode without directly copying the watermarked model parameters. 
However, Table \ref{table:exp_overview} shows that our proposed methods can successfully resist to distillation attack with great watermark identification effect that have the best watermark accuracy than all baselines.

\begin{table}[H]
\caption{Customization Ability on CIFAR-5. We have 5 different watermarked ResNet18 models on CIFAR-5.}
\label{table:per_cifar5}
\resizebox{\columnwidth}{!}{%
\def\arraystretch{1.6}
\begin{tabular}{ c|c|c|c|c|c } 

\textbf{CIFAR-5} & $M_{wm1}$ &  $M_{wm2}$ & $M_{wm3}$& $M_{wm4}$ &  $M_{wm5}$\\ \hline\hline

Main task accuracy. (\%) & 87.87 & 88.05 & 87.12 & 87.77 & 87.44 \\ 
\hline
wm det rate of $M_{d1}$ (\%) & \textbf{88.56} & 14.90 & 12.21 & 17.30 & 4.86\\ 
wm det rate of $M_{d2}$ (\%) & 10.48 & \textbf{86.17} & 14.34 & 11.64 & 11.63 \\ 
wm det rate of $M_{d3}$ (\%) & 16.15 & 12.36 & \textbf{85.06} & 5.88 & 0.72 \\ 
wm det rate of $M_{d4}$ (\%) & 16.47 & 6.76 & 5.88 & \textbf{87.52} & 0.41 \\ 
wm det rate of $M_{d5}$ (\%) & 2.15 & 7.62 & 2.17 & 0.61 & \textbf{86.53} 

\end{tabular}}
\end{table}
\begin{table}[H]
\caption{Customization Ability on CIFAR-10. We have 5 different watermarked ResNet18 models on CIFAR-10.}
\label{table:per_cifar10}
\resizebox{\columnwidth}{!}{%
\def\arraystretch{1.6}
\begin{tabular}{ c|c|c|c|c|c } 

\textbf{CIFAR-10} & $M_{wm1}$ &  $M_{wm2}$ & $M_{wm3}$& $M_{wm4}$ &  $M_{wm5}$\\ \hline\hline

Main task accuracy. (\%) & 94.67 & 94.80 & 94.82 & 94.46 & 94.67 \\ 
\hline
wm det rate of  $M_{d1}$ (\%) & \textbf{96.81} & 24.52 & 7.83 & 0.71 & 3.35 \\ 
wm det rate of  $M_{d2}$ (\%) & 18.35 & \textbf{98.86} & 0.24 & 2.99 & 0.35 \\ 
wm det rate of  $M_{d3}$ (\%) & 0.00 & 0.00 & \textbf{90.94} & 0.00 & 14.61 \\ 
wm det rate of  $M_{d4}$ (\%) & 0.73 & 3.41 & 0.40 & \textbf{98.54} & 0.57 \\ 
wm det rate of  $M_{d5}$ (\%) & 1.96 & 2.72 & 10.95 & 0.11 & \textbf{90.00} 

\end{tabular}}
\end{table}

\subsection {Customization Ability}
We train ResNet18 on the CIFAR-5, CIFAR-10, CIFAR-100, and Tiny ImageNet to evaluate the customization ability of our method. For each dataset, we generate 5 sets of watermarked model and detector, use {\emph{wm det rate.}} as metric. The {\emph{wm det rate.}} here indicates how confident the detector is in identifying the model. %The higher {\emph{wm det rate.}}, the more the detector considers the model to be its corresponding model.

\noindent \textbf{For  CIFAR-5 and CIFAR-10}, we set the signal strength factor $\gamma$ as 2 and 5,and set the length of signal vector $S$ to embed as the same number of labels in the respective dataset. %The results are shown in Table \ref{table:per_cifar5} and Table \ref{table:per_cifar10} respectively.

\noindent \textbf{For CIFAR-100}, we set the signal strength factor $\gamma$ as 5 and the length of signal vector $S$ to embed as 10. %The results is shown in Table \ref{table:per_cifar100}. 

\noindent \textbf{For Tiny ImageNet}, we set the signal strength factor $\gamma$ as 10 and the length of signal vector $S$ to embed as 10. %We present the results in Table \ref{table:per_tiny}.

\begin{table}[H]
\caption{Customization Ability on CIFAR-100. We have 5 different watermarked ResNet18 models on CIFAR-100.}
\label{table:per_cifar100}
\resizebox{\columnwidth}{!}{%
\def\arraystretch{1.6}
\begin{tabular}{ c|c|c|c|c|c } 

\textbf{CIFAR-100} & $M_{wm1}$ &  $M_{wm2}$ & $M_{wm3}$& $M_{wm4}$ &  $M_{wm5}$\\ \hline\hline

Main task accuracy. (\%) & 76.67 & 76.27 & 76.06 & 76.17 & 76.46\\ 
\hline
wm det rate of  $M_{d1}$ (\%) & \textbf{93.13} & 12.46 & 8.39 & 7.05 & 10.11 \\ 
wm det rate of  $M_{d2}$ (\%) & 9.77 & \textbf{91.59} & 8.53 & 6.32 & 13.77 \\ 
wm det rate of  $M_{d3}$ (\%) & 10.24 & 10.12 & \textbf{93.53} &  8.92 & 11.34 \\ 
wm det rate of  $M_{d4}$ (\%) & 8.14 & 7.44 & 15.32 & \textbf{91.22} & 9.53 \\ 
wm det rate of  $M_{d5}$ (\%) & 20.56 & 14.58 & 17.99 & 10.83 & \textbf{91.60} 

\end{tabular}%
}
\end{table}

\begin{table}[H]
\caption{Customization Ability on Tiny ImageNet. We have 5 different watermarked ResNet18 models on Tiny ImageNet.}
\label{table:per_tiny}
\resizebox{\columnwidth}{!}{%
\def\arraystretch{1.6}
\begin{tabular}{ c|c|c|c|c|c } 

\textbf{Tiny ImageNet} & $M_{wm1}$ &  $M_{wm2}$ & $M_{wm3}$& $M_{wm4}$ &  $M_{wm5}$\\ \hline\hline

Main task accuracy. (\%) & 67.00 & 67.30 & 67.24 & 67.11 & 67.16\\ 
\hline
wm det rate of $M_{d1}$ (\%) & \textbf{95.99} & 12.45 & 9.80 & 12.14 & 9.44\\ 
wm det rate of $M_{d2}$ (\%) & 19.88 & \textbf{96.04} & 21.14 & 12.34 & 8.43 \\ 
wm det rate of $M_{d3}$ (\%) & 17.91 & 18.82 & \textbf{95.11} & 5.71 & 11.54 \\ 
wm det rate of $M_{d4}$ (\%) & 23.91 & 12.38 & 16.87 & \textbf{91.57} & 11.56 \\ 
wm det rate of $M_{d5}$ (\%) & 8.02 & 18.88 & 15.47 & 17.34 & \textbf{93.73} 

\end{tabular}%
}
\end{table}

From Table \ref{table:per_cifar5},  \ref{table:per_cifar10},  \ref{table:per_cifar100}, and  \ref{table:per_tiny}, we can observe that their values on the diagonal are all very high, while the values of other parts are very low, all below 25 \%, which means that the recognition ability of each detector for its corresponding model is very high, and it will not mistake the model with other watermarks as its own. In addition, the results also indicate that our method has good customization ability regardless of the number of labels in the main task dataset.

\begin{table*}[ht]
\centering
\caption{Distillation Attack with Different \emph{cmps ratio}. We adopt 4 different student models, including ResNet18 (11,173,962 parameters), Mobilenet v2 (2,254,090 parameters), Shufflenet v2 (1,268,646 parameters), and PreResNet20 (272,282 parameters).}
\label{table:exp-diffstu}
\begin{adjustbox}{width=\textwidth}
\def\arraystretch{1.6}
\begin{tabular}{ccccccccccc}
\toprule

\multicolumn{1}{c}{ } & \multicolumn{2}{c}{\multirow{2}{*}{\textbf{ResNet18} (Teacher)}} & \multicolumn{2}{c}{\textbf{ResNet18} (Student)} & \multicolumn{2}{c}{\textbf{Mobilenet v2} (Student)} & \multicolumn{2}{c}{\textbf{Shufflenet v2} (Student)} & \multicolumn{2}{c}{\textbf{PreResNet20} (Student)}  \\
\cmidrule(rl){4-5} \cmidrule(rl){6-7} \cmidrule(rl){8-9} \cmidrule(rl){10-11} 
\multicolumn{1}{c}{ }& \multicolumn{2}{c}{~} & \multicolumn{2}{c}{cmps ratio. (\%): 1.00} & \multicolumn{2}{c}{cmps ratio. (\%): 4.95} & \multicolumn{2}{c}{cmps ratio. (\%): 8.81} & \multicolumn{2}{c}{cmps ratio. (\%): 41.03}\\
\cmidrule(rl){2-3} \cmidrule(rl){4-5} \cmidrule(rl){6-7} \cmidrule(rl){8-9} \cmidrule(rl){10-11} 
\textbf{Method} & {main acc.(\%)} & {wm acc.(\%)}  & {main acc.(\%)} & {wm acc.(\%)} & {main acc.(\%)} & {wm acc.(\%)} & {main acc.(\%)} & {wm acc.(\%)} & {main acc.(\%)} & {wm acc.(\%)} \\
\hline
\midrule
\texttt{WNN}& 90.67 & 100.00 & 92.24 & 8.00 & 90.12 & 8.52 & 89.11 & 11.00 & 89.03 & 9.10\\
\texttt{DeepIPR}& 92.11 & 100.00 & 90.14 & 0.00 & 91.32 & 0.00 & 90.95 & 0.00 & 90.42 & 0.00\\
\texttt{PA}& 94.02 & 100.00 & 92.33 & 0.00 & 92.03 & 0.00 & 90.21 & 0.00 & 90.51 & 0.00\\
\texttt{RIGA}& 94.43 & 100.00 & 94.12 & 0.00 & 92.76 & 0.00 & 91.17 & 0.00 & 90.93 & 0.00\\
\texttt{DAWN}& 79.17 & 99.60 & 78.11 & 86.20 & 77.31 & 86.19 & 70.54 & 75.34 & 68.54 & 72.47\\
\texttt{CosWM}& 95.14 & 88.28 & 95.15 & 83.54 & 92.21 & 83.03 & 92.71 & 84.89 & 92.66 & 83.92 \\
\hline
\textbf{USP (Ours)}& 93.75 & 99.56 & 94.92 & \textbf{99.27} & 92.06 & \textbf{99.08} & 90.91 & \textbf{99.34} & 90.22 & \textbf{98.26}  \\
\textbf{CSP (Ours)}& 94.71 & 98.68 & 90.05 & \textbf{91.86} & 92.12 & \textbf{96.21} & 92.86 & \textbf{94.94} & 92.74 & \textbf{92.24} \\
\bottomrule
\end{tabular}
\end{adjustbox}
\end{table*}

\begin{table}[H]
\caption{Customization of different watermarking method on CIFAR-10. \emph{Identified rate} indicates the average rate that the correct watermark is successfully recognized (true positive) . \emph{Misidentified rate} indicates the average rate that the wrong watermark is misidentified (false positive) .}
\label{table:exp-per-cmp}
\resizebox{\columnwidth}{!}{%
\def\arraystretch{1.6}
\begin{tabular}{ccccccccc}

{} & \textbf{Identified rate (\%)} & \textbf{Misidentified rate (\%)}\\
\hline
\midrule
\texttt{WNN} (USENIX'2018)& 100.00& 10.34 \\
\texttt{DeepIPR} (NIPS'2019)& 100.00& 6.93 \\
\texttt{PA} (NIPS'2020)& 100.00& 5.49 \\
\texttt{RIGA} (WWW'2021)& 100.00 & \underline{\textit{25.00}}\\
\texttt{DAWN} (MM'2021)& \underline{\textit{88.80}} & 0.00\\
\texttt{CosWM} (AAAI'2022)& \underline{\textit{85.57}}& \underline{\textit{57.34}} \\
\hline
%\textbf{USP (Ours)} & 96.60 & \underline{\textit{93.68}}\\
\textbf{CSP (Ours)} & \textbf{99.16}& \textbf{0.97}\\
\end{tabular}%
}
\end{table}

To see how our customization ability compares to existing methods, we generate a total of 3 watermarking models on CIFAR-10 for each baseline method and use \emph{True identified rate} and \emph{Misidentified rate} to evaluate the customization ability of each method. For watermarking methods that need to assign a bit string or a vector as watermark embedding, such as feature-based methods, we set the similarities of the bit strings or vectors of the three watermarked models to each other to be 50\%. For \texttt{CosWM}, we set 3 different target class and random unit projection vector for each watermarked model. The experimental results are presented in Table \ref{table:exp-per-cmp}. \emph{True identified rate} represents the rate at which the correct watermark is successfully recognized, for example, $wm_{1}$ detected from $M_{wm_{1}}$, and \emph{Misidentified rate} represents the rate at which the wrong watermark is misidentified. %, for example, average of the watermark accuracy of $wm_{2}$ detected from $M_{wm_{1}}$ and the watermark accuracy of $wm_{3}$ detected from $M_{wm_{1}}$.
From Table \ref{table:exp-per-cmp}, we can find that \texttt{DAWN} seems to work well on the misidentified rate. Because \texttt{DAWN} directly uses a subset of the main task dataset as the trigger set, the same image will have completely different labels in the training data of different watermarked models.%, making the models not misidentify each other. 
However, \texttt{DAWN} is the worst at recognizing its own watermark.%, which is also the cost of directly using the main task dataset as a trigger set.
\emph{CSP} achieves 0.97\% of misidentified rate which is much lower than most of the methods, and it performs well in true identified rate which is as high as 99.16\%. The results show that, compared with baselines, \emph{CSP} achieves a better balance between the self-recognition effect and the probability of misidentification.%, indicating that it has a great customization ability.

\subsection {Robustness against Knowledge Distillation Attack with Different Compression Rate}

From Table \ref{table:exp_overview}, we know most of the watermarking methods can not resist to knowledge distillation attack, except \texttt{CosWM} and our proposed approach. To better understand the defense power of these watermarking methods against distillation attacks, we adopt 4 different model architectures to be the student models, including ResNet18, Mobilenet v2, Shufflenet v2, and PreResNet20, on CIFAR-10. We set the training(distilling) epochs as 160, learning rate as 0.1, and distillation temperature as 4.

\noindent The results are summarized in Table \ref{table:exp-diffstu}. We can find that our methods always achieve the best watermark accuracy. On the student ResNet18 model, our methods have 99.27\% and 91.86\% watermark accuracy, and on the smallest student model, PreResNet20, our methods also have 98.26\% and 92.24\% watermark accuracy. It indicates 
%that either the student model, which is as large as the teacher model, or the student model, which is compressed by about 41 times, the watermark embedded by our methods can be well preserved in. Thus, it can be seen from the above results 
that our methods are robust to distillation attacks that not limited by the model compression size.

\section{Conclusion}
In this paper, we design a deep neural network watermarking framework \emph{Unified Soft-label Perturbation (USP)}, which has a detector paired with the model to be watermarked, to defend against distillation attacks. We further propose \emph{Customized Soft-label Perturbation (CSP)} to embed the watermark into the output via perturbing the prediction probability distribution. Experiments show that our proposed approaches successfully resist different kinds of watermarking removal attacks, and outperform the other state-of-the-art baselines in tougher attack settings.

%%
%% The acknowledgments section is defined using the "acks" environment
%% (and NOT an unnumbered section). This ensures the proper
%% identification of the section in the article metadata, and the
%% consistent spelling of the heading.
%\begin{acks}
%To Robert, for the bagels and explaining CMYK and color spaces.
%\end{acks}

%%
%% The next two lines define the bibliography style to be used, and
%% the bibliography file.
\bibliographystyle{ACM-Reference-Format}
\bibliography{main}

%%
%% If your work has an appendix, this is the place to put it.
\appendix

\section {Details of Evaluation Setup}
\subsection {Environments.}

We implement our framework, \texttt{RIGA} and \texttt{CosWM} in PyTorch 1.3. For \texttt{WNN, DeepIPR, PA,} and \texttt{DAWN}, we adopt their source code provided on the github. \texttt{WNN} is implemented in PyTorch 0.4.1. \texttt{DeepIPR} and \texttt{PA} are implemented in PyTorch 1.3. DAWN is implemented in PyTorch 1.4. All the experiments are conducted with Intel (R) Core (TM) i9-9900K CPU @ 3.60GHz, NVIDIA 2080Ti, and Ubuntu 18.04.6.

\subsection{Datasets.} 
We evaluate our method on four different datasets: CIFAR-5, CIFAR-10, CIFAR100, and Tiny ImageNet datasets. 

\noindent \textbf{CIFAR-10} has $10$ labels with $50,000$ training and $10,000$ testing images. For each label, there are $5,000$ training and $1,000$ testing images. The resolution of each color image is $32\times32$.

\noindent  \textbf{CIFAR-5} is a subset of CIFAR-10 that consists of data with label 0 to 4 in CIFAR-10. Thus, each lable has $5,000$ training and $1,000$ testing images as CIFAR-10, and the resolution of image is also $32\times32$.

\noindent  \textbf{CIFAR-100} has $100$ classes, and each class obtains $500$ training images and $100$ testing images. The resolution of each color image is also $32\times32$ as as CIFAR-10.

\noindent  \textbf{Tiny ImageNet} contains $100,000$ color images of $200$ classes. There are $500$ training images, $50$ validation images and $50$ test images for each class. The images size is $64\times64$.

\subsection{Experiment Setting.}

\noindent \textbf{Our method.} We train ResNet18 on four datasets. On CIFAR-5 and CIFAR-10, we set batch size as $512$. On CIFAR-100 and Tiny ImageNet, we set batch size as $256$. On CIFAR-5 and CIFAR-10, the detector learning rate decreases by timing $0.1$ in epoch $140$. On CIFAR-100 and Tiny ImageNet, the detector learning rate decreases by timing $0.1$ in epoch $160$.

\noindent \textbf{Baselines.} For \texttt{CosWM}, the learning rate starts from $0.1$ and decreases by timing $0.1$ in epoch $80$, $120$. The optimizer is SGD with Nesterov momentum setting momentum as $0.9$ and weight decay as $5\times{10}^{-4}$. For \texttt{RIGA}, we set training epochs as $160$ and batch size as $128$, adopt Adam as optimizer, and make the learning rate starts from $0.001$ and decreases by timing $0.1$ in epoch $80$, $120$.

\begin{table}[ht]
\caption{Customization ability of CSP on CIFAR-10.}
\label{table:exp-per-csp}
\resizebox{\columnwidth}{!}{%
\def\arraystretch{1.6}
\begin{tabular}{cccc}

\# Models & \textbf{Identified rate (\%)} & \textbf{Misidentified rate (\%)} & \textbf{F1 score (\%)}\\
\hline
\midrule
3& 99.168 & 0.977 & 99.982\\
5& 98.604 & 0.867 & 99.210\\
7& 98.892 & 11.194 & 76.142\\
10& 96.114& 10.510 & 76.031\\
13& 96.884 & 9.777 & 76.418\\
15& 95.335& 8.397 & 74.005\\
17& 95.058 & 8.507 & 70.124\\
20& 94.977 & 8.875 & 69.412\\

\end{tabular}%
}
\end{table}

\section{Customization ability of CSP}
Then we further evaluate the customization ability of \emph{CPS} with a larger number of models and the results are in Table \ref{table:exp-per-csp}. We vary the number of models from 3 to 20. We can see from the results that the average misidentified rate will increase as the number of models increases, because the total number of labels in the dataset is fixed. When the number of models increases, the range of disturbances that the watermark needs to add to the output will become larger, making The locations that add perturbations to each other have an increased chance of overlapping, which in turn affects the misidentified rate. Even so, compared with other methods, the average misidentified rate of \emph{CSP} is still low and seems to converge at around 8\%~10\%, and the average identified rate and the average remains very high, still more than 94\%.

\section {Re-embed Threshold}

In order to find the most suitable re-embed threshold $\tau$, we evaluate the performance of \emph{CSP} under different $\tau$ settings, and also apply fine-tuning attack and pruning attack with re-training to know the robustness of each model. We vary $\tau$ in [75, 80, 85, 90, 95]. For fine-tuning attack, we fine-tune the models with 100 epochs and a learning rate of 0.01. For pruning attack with re-training, we prune off 80\% parameters, setting the re-training epoch and learning rate as 100 and 0.01.  The results are summarized in Table \ref{table:re-emb}. 

From the experimental results, it can be found that $\tau$ has a large impact on the robustness of the detector. If we set lower $\tau$, the model focuses on the main task during the overall fine-tuning step, so the watermark is sacrificed to obtain better main task performance. When we apply fine-tuning attack and pruning attack with re-training, the watermark accuracy will drop soon in early epochs. 
However, if set higher $\tau$, the accuracy of the watermark is excessively considered, the model will continue to deepen the watermark. In addition to reducing the accuracy on the main task, the total computation time will become longer, and the robustness of the detector will also deteriorate. Because the detector will over-fit on the well-engraved watermark training data and has never seen detection training data with any augmentation. Thus, we can find that the watermarked models with high \ref{table:re-emb} have relatively poor watermark accuracy after attacking.

\begin{table}[H]
\caption{Re-embed Threshold $\tau$. To find the most suitable threshold, we vary the value of $\tau$ and evaluate the robustness of each setting.}

\label{table:re-emb}
\resizebox{\columnwidth}{!}{%
\def\arraystretch{1.6}
\begin{tabular}{ c|c|c|c|c|c|c } 

\multicolumn{2}{c|}{$\tau$} &  75.00  & 80.00 & 85.00 & 90.00 & 95.00 \\
\hline
\hline
\multicolumn{2}{c|}{main acc. (\%) of \textbf{Unwatermarked model}} & \multicolumn{5}{c}{95.30}\\
\hline
\hline
\multirow{2}{*}{\textbf{Orignal}} & main acc. (\%) & 94.67 & 94.66 & 94.62 & 94.55 & 94.55\\
%\cline{2-7}
~ & wm acc. (\%) & 97.50 & 98.11 & 98.16 & 98.41 & 98.41\\
\hline
\textbf{Fine-tuning} & main acc. (\%) & 94.70 & 94.71 & 94.61 & 94.67 & 94.65\\
%\cline{2-7}
\textbf{attack} & wm acc. (\%) & 64.48 & 62.03 & \textbf{88.72} & 52.32 & 50.99\\
\hline
\textbf{Pruning} & main acc. (\%) & 94.24 & 94.12 & 94.19 & 94.23 & 94.11\\
%\cline{2-7}
\textbf{attack} & wm acc. (\%) & 69.03 & 71.25 & \textbf{87.40} & 58.87 & 56.32 \\

\end{tabular}%
}
\end{table}

\section {Fine-tuning Attack with Different Learning Rate}

According to our preliminary analysis, some watermarking methods can resist the fine-tuning attack only when the learning rate is small. 
%In Tabel \ref{table:exp_overview}, our proposed approaches outperform under fine-tuning attack with larger learning rate. 
Thus, we apply fine-tuning attack with small learning rate, 0.001, to our approaches here to evaluate our robustness to different fine-tuning learning rate settings, and the results are summarized in Table \ref{table:exp-difflr}.% From the experimental results, the watermark accuracy of our approaches remains high after fine-tuning 100 epochs, whether under the setting of high learning rate or low learning rate. It represents our strong robustness to fine-tuning attacks.

\begin{table}[H]
\caption{Fine-tuning attacks with different learning rate on CIFAR-10. We take the result with the highest main acc.}
\label{table:exp-difflr}
\resizebox{\columnwidth}{!}{%
\def\arraystretch{1.6}
\begin{tabular}{ccccc}

\multicolumn{1}{c}{ } & \multicolumn{2}{c}{\textbf{lr = 0.001}} & \multicolumn{2}{c}{\textbf{lr = 0.01}}\\
\cmidrule(rl){2-3} \cmidrule(rl){4-5}

 & {main acc. (\%)} & {wm acc.(\%)} & {main acc.(\%)} & {wm acc.(\%)}\\
\hline
\midrule
\texttt{WNN} (USENIX'18)&90.72& 100.00& 92.43 & \underline{\textit{52.31}} \\
\texttt{DeepIPR} (NIPS'19)& 92.40 & 100.00& 92.55 & \underline{\textit{70.48}} \\
\texttt{PA} (NIPS'20)& 94.07& 100.00& 92.26 & \underline{\textit{70.73}} \\
\texttt{RIGA} (WWW'21)& 94.43 & 100.00 & 93.34 & 99.24 \\
\texttt{DAWN} (MM'21)& 79.20 & \underline{\textit{8.80}} & 80.34 & \underline{\textit{0.00}}\\
\texttt{CosWM} (AAAI'22)& 94.95& 88.35& 95.00 & 82.97 \\
\hline
\textbf{USP (Ours)}& 93.38 & \textbf{98.60} & 93.84 & \textbf{86.28} \\
\textbf{CSP (Ours)}& 94.64& \textbf{98.61}& 94.60 & \textbf{88.72}\\
\end{tabular}%
}
\end{table}

\begin{table}[ht]
\caption{Ablation Experiment of ${L}_{KLD}$ and Detector.}
\label{table:exp-abl}
\resizebox{\columnwidth}{!}{%
\def\arraystretch{1.6}
\begin{tabular}{cccc}

 & \textbf{main acc. (\%)} & \textbf{wm acc. (\%)} & \textbf{${L}_{KLD}$}\\
\hline
\midrule
With ${L}_{KLD}$ and detector & 94.947 & 98.886 & 2.914 \\
Only with detector & 94.976 & \underline{\textit{66.233}} & - \\
Only with ${L}_{KLD}$ & 94.036 & - & \textbf{3.581} \\
Unwatermarked model ${M}_{2}$& 93.989 & - & \textbf{2.105} \\

\end{tabular}%
}
\end{table}

\section {Ablation Experiment - kl divergence loss and Detector}

To know how the kl divergence loss (${L}_{KLD}$) and the detector affect our methods, we conduct an ablation experiment on \emph{USP}. Since there is a customized perturbation adding mechanism in \emph{CSP}, it is not easy to see the direct impact of ${L}_{KLD}$ on the method. The results are showed in Table \ref{table:exp-abl}. First of all, if only the detector is used, the watermark accuracy becomes very low. We consider the reason is when the model updates its parameters without the objective of increasing ${L}_{KLD}$ makes the deviation in the output is not enough, then resulting in the poor effect of the detector. If only ${L}_{KLD}$ is used and no detector is used, there is no standard for judging whether the watermark exists. Here, we also train an unwatermarked model ${M}_{2}$ with a comparable main task accuracy. From Table \ref{table:exp-abl}, we can find the ${L}_{KLD}$ of ${M}_{2}$ is also large, so just using ${L}_{KLD}$ is also not enough.

\end{document}